%% file: paper.tex
\documentclass[runningheads]{llncs}

\usepackage[T1]{fontenc}
\usepackage{amsmath}
\usepackage{listings}
\usepackage{microtype}
\usepackage{hyperref}
\usepackage{graphicx}
\usepackage{setspace}
\usepackage{placeins}
\usepackage{tikz}
\usetikzlibrary{positioning, shapes.misc, shapes.geometric, arrows.meta, backgrounds, bending}
\usetikzlibrary{fit}

\definecolor{codegreen}{rgb}{0,0.6,0}
\definecolor{codegray}{rgb}{0.5,0.5,0.5}
\definecolor{codepurple}{rgb}{0.58,0,0.82}
\definecolor{backcolour}{rgb}{0.95,0.95,0.92}

\lstdefinestyle{mystyle}{
    commentstyle=\color{codegreen},
    keywordstyle=\color{magenta},
    numberstyle=\tiny\color{codegray},
    stringstyle=\color{codepurple},
    basicstyle=\ttfamily\footnotesize,
    breakatwhitespace=false,
    breaklines=true,
    captionpos=b,
    keepspaces=true,
    numbers=left,
    numbersep=5pt,
    showspaces=false,
    showstringspaces=false,
    showtabs=false,
    tabsize=2
}
\begin{document}

\title{Silent Data Corruption Protection\\through Efficient Task Replication}

\titlerunning{Silent Data Corruption Protection through Efficient Task Replication}

\author{Mia Reitz \and Claudia Fohry
}

\authorrunning{M. Reitz et al.}

\institute{\textit{Research Group Programming Languages / Methodologies}\\ \textit{University of Kassel}\\ Kassel, Germany \\
\email{\{ mia.reitz~\textbar~fohry \}@uni-kassel.de}
}

\maketitle

\begin{abstract}
\input{00abstract.tex}
\end{abstract}

\input{01intro.tex}
\FloatBarrier
\input{02relatedwork.tex}
\FloatBarrier
\input{03background.tex}
\FloatBarrier
\input{04design.tex}
\FloatBarrier
\input{05experiments.tex}
\FloatBarrier
\input{06extension.tex}
\input{07conclusions.tex}
\FloatBarrier

\begin{credits}
\subsubsection{\ackname} 
This research was funded by the Deutsche Forschungsgemeinschaft (DFG, German
Research Foundation) under project number 512078735.
The authors gratefully acknowledge the computing time provided to them on the Goethe-NHR cluster at the Frankfurt Center for Scientific Computing.

\subsubsection{\discintname}
The authors have no competing interests to declare that are
relevant to the content of this article.
\end{credits}

\bibliographystyle{splncs03_unsrt}
\bibliography{bibo}

\end{document}

%% file: 00abstract.tex
The trend of increasing cluster sizes of supercomputers leads to a growing susceptibility to Silent Data Corruption (SDC) that can invalidate program results.
A common strategy for SDC protection is replication, where the computation is repeated, and the correct result is determined as the one that is the same in at least two different computations.
Applying replication to Asynchronous Many-Task (AMT) runtimes on clusters is challenging due to dynamic task spawning and work stealing, which complicate the identification of replicated tasks.

To address the challenge, this paper introduces a novel replication scheme that detects and corrects SDCs for nested fork-join programs.
Briefly stated, our approach replicates the computation and records the task tree.
Upon a mismatch in the final result, it traverses the tree top-down to identify all corrupted tasks that could have impacted the final result.
Recovery is then performed by recomputing these tasks, while the results of correct child tasks are reused.

We demonstrate our implementation within a variant of the Itoyori cluster AMT runtime.
Our experimental results suggest that the time to identify and reprocess the affected tasks is negligible.
The paper concludes by discussing the adaptability of our scheme to tasks that cooperate through futures.

%% file: 01intro.tex
\section{Introduction}\label{sec:intro}

The trend of increasing cluster sizes in supercomputers leads to a growing susceptibility to hardware faults.
Among the most challenging of these are Silent Data Corruptions~(SDCs), which can invalidate program results without causing a program crash~\cite{FTBuch}.
Consequently, SDC protection is essential for applications running on supercomputers.
A common strategy is replication, which is applicable to both SDC detection and correction.
When deployed with one replica, replication allows to detect SDCs, and when deployed with multiple replicas, it also allows to correct them via a majority vote.
Replication has a high computational overhead, especially when used with multiple replicas.

An established approach for programming parallel applications is Asynchronous Many-Tasking~(AMT), where the dynamic nature of computations presents a challenge for implementing efficient SDC protection.
AMT programs decompose a computation into a large number of fine-grained tasks that are processed by \emph{worker} processes.
Load balancing is typically achieved through work stealing, where idle workers take tasks from busy ones.
We specifically consider nested fork-join~(NFJ), in which each task can create child tasks via a \texttt{spawn} operation, and later waits for the results of all of its children.
The tasks and their dependencies give rise to a task tree.

This paper proposes a novel replication scheme that detects SDCs for NFJ programs and corrects them in a more efficient way than sketched above.

We assume that failures only occur within the execution of tasks whereas the memory, the runtime system, and the communication layer are reliable.
Our approach runs the computation twice by independently launching two identical root tasks: an original one and a \textit{twin}.
During execution, it logs the task trees and the results of both computations.
If the final results of the computations differ, an SDC has occurred.
Recovery is then started with a top-down traversal of the task trees,
which identifies all corrupted tasks that could have impacted the final result.
Afterwards, the computation is re-run, whereby the identified tasks are recomputed and the results of their healthy child tasks are reused.

The main contributions of this paper are:
\begin{itemize}
    \item A novel replication-based scheme for SDC detection and correction for NFJ programs.
    \item An implementation and evaluation of this scheme within a variant of the Itoyori~\cite{MiaFutureSideEffects} runtime that supports NFJ programs.
Preliminary experimental results demonstrate correctness and suggest that the time required to identify and reprocess affected tasks is negligible.
    \item A discussion of the adaptability of our scheme to tasks that cooperate through futures.
\end{itemize}

The remainder of this paper is organized as follows.
Next, Section~\ref{sec:related_work} surveys related work.
Section~\ref{sec:background} provides background on NFJ programs and their realization in the Itoyori variant.
Then, Section~\ref{sec:design} details our fault tolerance scheme.
Section~\ref{sec:experiments} presents and discusses experimental results, whereas
the adaptability of the scheme to tasks with futures is discussed in Section~\ref{sec:extension}.
Finally, Section~\ref{sec:conclusions} concludes the paper and outlines future work.

%% file: 02relatedwork.tex
\section{Related Work}\label{sec:related_work}

Fault tolerance strategies typically address two distinct failure types: fail-stop errors, where processes crash, and SDCs, where components produce incorrect results.
Significant research has addressed fail-stop failures in Asynchronous Many-Task~(AMT) runtimes with task-level checkpointing~\cite{JonasIncFTGLBJournal19,MiaNFJCheckpointing,MiaFuturesCheckpointing,JonasIJNC22} and supervision~\cite{JonasIJNC22,KrishnaNestedFJournal} techniques.
Another approach is proposed in the HOPE execution model~\cite{HOPE}, where workers run the same program and exchange results to omit tasks.
In a similar way, our SDC protection scheme omits correct tasks during the reprocessing phase.

A common approach for SDC detection is replication, where a computation is executed multiple times and the results are compared.
In traditional replication, the computation is run three times such that the failures can be detected and corrected via a majority vote, but at an overhead of a factor of three~\cite{originalReplication}.
Our scheme adopts a more efficient approach, which deterministically detects and corrects failures with essentially a factor of two overhead.

SDC detection may also be realized with user-provided validation functions~\cite{ftx1}, and
SDC protection may be realized with Algorithm-Based Fault Tolerance~(ABFT), in which checksums and other mathematical invariants are directly embedded into an algorithm~\cite{ABFTBos,ABFT,ABFTOrig}.
However, ABFT has limited generality, and is often inapplicable to irregular or complex applications.
In contrast, our replication-based method treats tasks as black boxes and provides a universal protection mechanism for NFJ-type AMT runtimes.

Other recent work on SDC protection for AMT runtimes focuses
exclusively on SDC correction, assuming the availability of an external detection mechanism, and performs a localized re-execution of corrupted tasks~\cite{SoftKrishna}.
Also, application-specific data redundancy can be used to reconstruct corrupted data, falling back to partial replication when reconstruction fails~\cite{DoubtAndRedundancySoft}.

%% file: 03background.tex
\section{Background}\label{sec:background}

\subsection{Nested Fork Join~(NFJ)}
\begin{lstlisting}[label=lst:nfj_example,caption={Pseudocode of a naive Fibonacci implementation in \textbf{NFJ}.},float=t,frame=tlrb, xleftmargin=1em, basicstyle=\ttfamily\small, numbers=left, captionpos=b, morekeywords={spawn,sync}]
int result = fib(n);

int fib(int n) {
  if (n < 2) return 1;
  int a = spawn fib(n - 1);
  int b = spawn fib(n - 2);
  sync;
  return a + b;
}
\end{lstlisting}
The NFJ model is a natural fit for divide-and-conquer algorithms.
As illustrated by a naive Fibonacci computation in Listing~\ref{lst:nfj_example}, the computation starts with a single task.
Then, each parent task can \texttt{spawn} one or more child tasks to compute sub-problems.
Later, the parent task invokes a \texttt{sync} operation, which suspends its execution until all previously spawned children have returned their results.
A task may \texttt{sync} multiple times, e.g., between and after the two \texttt{spawn}s in the listing, to first wait for part of the results.
Eventually, it must wait for all results.
The parent-child relationship gives rise to a task tree, in which
the root task initiates the computation, and results are passed up the tree, allowing the root task to calculate the final result.
We assume that tasks do not have side effects.

\subsection{ItoyoriFBC Runtime}
\begin{lstlisting}[label=lst:fbcexample,caption={Pseudocode of a naive Fibonacci implementation using futures.},float=t,frame=tlrb, xleftmargin=1em, basicstyle=\ttfamily\small, numbers=left, captionpos=b, morekeywords={spawn,touch,future}]
int result = fib(n);

int fib(int n) {
  if (n < 2) return 1;
  future<int> a = spawn fib(n - 1);
  future<int> b = spawn fib(n - 2);
  return a.touch() + b.touch();
}
\end{lstlisting}
Itoyori~\cite{itoyori} is a C++-based AMT runtime.
We use a variant, called ItoyoriFBC~\cite{MiaFutureSideEffects}, that allows tasks to communicate through futures.
A future is a placeholder for a result that will be computed eventually by a task.
When a task is spawn, it immediately returns a future.
Other tasks can wait for and retrieve the result held in a future by \textit{touching} it.
If the result is already available, it is returned immediately;
otherwise, the touching task is suspended until the value has been computed.
As illustrated in Listing~\ref{lst:fbcexample}, futures can be used for coding NFJ programs.

In ItoyoriFBC, each worker is an MPI process. The workers communicate through one-sided MPI functions, managing a \emph{global address space}.
Futures are stored in the memory of the worker that created the future, even if the future-filling task has been stolen.
When a task on a different worker touches the future, the runtime uses the global address of the future to automatically fetch the required data.

ItoyoriFBC employs work stealing.
Thereby each worker maintains its own local queue of tasks.
In a task spawn, it puts the continuation of the parent task into this queue and branches into the child.
After finishing a task, it takes the next one from the queue. 
When the queue is empty, the worker acts as a \textit{thief} and attempts to steal a task from the queue of a randomly chosen \textit{victim} worker.
The runtime uses a synchronized data structure for the queues to prevent concurrent accesses to the same entry.

%% file: 04design.tex
\section{Fault Tolerance Scheme}\label{sec:design}

Our SDC protection scheme comprises three phases:
\begin{enumerate}
    \item \textbf{Tracking phase}: During the original and twin computations, the runtime records all tasks and their results in a task tree.
    \item \textbf{Traversal phase} (if needed): After the completion of both computations, their final results are compared.
If they differ, the saved task trees are traversed simultaneously to identify all corrupted tasks with impact on the final result.
    \item \textbf{Reprocessing phase} (if needed): Finally, the computation is re-run, whereby the identified tasks are reprocessed and all others are omitted.
\end{enumerate}
The following subsections detail the phases.

\subsection{Tracking Phase}

In both the original and twin computations, we record the dependency from a parent task~A to a child task~B.
We do this by expanding the future object that task~A generates for the result of task~B by a list of pointers to the futures of the child tasks of~B.

\subsection{Traversal Phase}\label{subsec:frontier}

\input{04_fig1.tex}

If the results of the original and twin computations agree, the program ends.
Otherwise, the traversal phase begins.

In this phase, the runtime inspects the relevant parts of the two task trees simultaneously in a depth-first order to classify tasks as corrupted or reusable.
In general, the traversal algorithm is as follows:

\begin{enumerate}
    \item Starting at the root, mark the current task (or actually its associated future object) as corrupted.
    \item Compare the result of each child~Z across the two computations.
      If the two values agree, do not continue the traversal along this path.
      Otherwise, apply the algorithm recursively to~Z.
\end{enumerate}

In the case that the number of children of a node disagrees between the original and twin computations, the program aborts with an error message.
Otherwise, upon termination, the algorithm has marked all corrupted tasks that could have contributed to the final result being wrong.

Note that the algorithm also works for multiple SDCs.
The traversal is triggered by the mismatch of the final result and will find every corrupted branch in the task tree that could have impacted the final result.

Figure~\ref{fig:corruption_frontier} depicts an example outcome of this phase:
    Dashed edges such as the one from~A to~B indicate that~B was spawn by~A.
    Solid edges such as the one from~B to~A indicate that the result of~B is read by~A.
Thus, both types of edges mark task dependencies.
In the example, SDCs originate in tasks~H and~F, corrupting their results.
    The corruption of~H propagates up and down the tree through result return~(to~D,~B, and~A) and parameter passing~(to~J and~E).
    C~is marked in orange to indicate that it is infected by the corruption of~F, but still returns a correct result to~A.
    Only the tasks within the red dashed area are reprocessed, while the results of their correct direct dependencies (including~C and the green tasks~G,~I, and~K) are reused.
Note that results of unconnected tasks, such as the blue task (L) are indirectly reused.

\subsection{Reprocessing Phase}
Once the corrupted tasks have been identified, the runtime restarts the computation by spawning the root task.
Whenever a task spawns a child task, it inspects the corresponding node in the original task tree.
If this node is marked as healthy, the spawning is suppressed and the recorded result is reused instead.
This phase needs to be repeated in the case when a task result differs from both of its previous two values, which is recognized during reprocessing. 

\subsection{Analysis}\label{subsec:analysis}
The overhead of our scheme in a failure-free execution is dominated by the overhead for the twin computation.
The additional overhead for the tracking of the task tree is negligible, as it involves only a single low-cost memory write per \texttt{spawn}.

To quantify the recovery cost after a single SDC, we analyze the expected number of tasks that must be reprocessed.
For simplicity, we assume that all tasks have the same running time, and that each of them is hit by the SDC with the same probability.
Furthermore, we consider a perfect binary task tree of height~\(h\), which contains a total of \(2^{h+1}-1\) tasks.
Recall that an SDC at task~\(X\) at height~\(d\) may cause corruption of:
\begin{enumerate}
    \item All $d+1$~tasks on the path from~\(X\) to the root. If only part of these tasks is affected, the failure does not cause any costs.
    \item The task subtree rooted in~\(X\).
This subtree contains \(2^{h-d+1}-1\)~tasks, and, in the worst case, all of them may be corrupted.
\end{enumerate}
Thus, the total number of tasks to be reprocessed,~\(N(d)\), is the sum of the tasks on the path and in the subtree:
\[ N(d) = (d+1) + (2^{h-d+1}-1) = d + 2^{h-d+1} \]

The number of tasks at depth~\(d\) is~\(2^d\), and the probability~\(P(d)\) of an SDC occurring at depth~\(d\) equals the ratio of tasks at that level to the total number of tasks:
\[ P(d) = \frac{2^d}{2^{h+1}-1} \]
The expected number of reprocessed tasks, \(E[C]\), is the sum of~\(N(d)\) over all possible depths, weighted by the probability of an SDC occurring at this depth:
\begin{align*}
E[C] = \sum_{d=0}^{h} P(d) \cdot N(d) &= \frac{1}{2^{h+1}-1} \sum_{d=0}^{h} 2^d (d + 2^{h-d+1}) \\
&= \frac{1}{2^{h+1}-1} \left(\sum_{d=0}^{h} d\,2^d + \sum_{d=0}^{h} 2^{h+1}\right) \\
&\text{$\ldots$simplifying the arithmetic-geometric sum$\ldots$} \\
&= 2h + \frac{2h + 2}{2^{h+1} - 1} \approx 2h
\end{align*}
This result shows that the expected number of reprocessed tasks is small. It scales linearly with the tree height ($\approx 2h$), whereas the total number of tasks grows exponentially.

\subsection{Implementation}
We implemented the scheme in ItoyoriFBC.
For the purposes of this study, we chose to execute the traversal phase on a single worker, to be able to measure its duration.
While this simplifies the implementation and allows for precise time measurements, it represents a scalability bottleneck.
Therefore, future work should parallelize the implementation of this phase, as well.

%% file: 04_fig1.tex
\begin{figure}[t]
\centering
\resizebox{\columnwidth}{!}{\begin{tikzpicture}[
    node distance=0.8cm and 1.2cm,
    task/.style={
        draw,
        rectangle,
        rounded corners,
        minimum size=8mm,
        font=\sffamily
    },
    corrupted/.style={
        task,
        fill=red!30,
        thick
    },
    infected/.style={
        task,
        fill=orange!50,
        thick
    },
    correct/.style={
        task,
        fill=green!20
    },
    frontier/.style={
        task,
        fill=blue!30,
    },
    sdc_source/.style={
        corrupted,
        star,
        star points=7,
        star point ratio=0.5,
        minimum size=7mm,
        line width=0.6mm
    },
    arrow/.style={
        -Latex,
        thick
    },
    label/.style={
        font=\sffamily\small
    }
]

\node[corrupted] (A) {A};

\node[corrupted, below left=of A] (B) {B};
\node[infected, below right=of A] (C) {C};

\node[corrupted, below left=of B] (D) {D};
\node[corrupted, below right=of B] (E) {E};
\node[sdc_source, below right=of C] (F) {F};

\node[correct, below left=of D] (G) {G};
\node[sdc_source, below right=of D] (H) {H};
\node[correct, below right=of E] (I) {I};

\node[corrupted, below left=of H] (J) {J};
\node[correct, below right=of H] (K) {K};
\node[frontier, below right=of I] (L) {L};

\draw[arrow] (B) -- (A);
\draw[arrow] (C) -- (A);

\draw[arrow] (D) -- (B);
\draw[arrow] (E) -- (B);

\draw[arrow] (F) -- (C);

\draw[arrow] (H) -- (D);
\draw[arrow] (I) -- (E);
\draw[arrow] (G) -- (D);

\draw[arrow] (J) -- (H);
\draw[arrow] (K) -- (H);
\draw[arrow] (L) -- (I);

\draw[arrow,dashed] (A) to[bend right=20] (B);
\draw[arrow,dashed] (A) to[bend right=20] (C);

\draw[arrow,dashed] (B) to[bend right=20] (D);
\draw[arrow,dashed] (B) to[bend right=20] (E);

\draw[arrow,dashed] (C) to[bend right=20] (F);

\draw[arrow,dashed] (D) to[bend right=20] (G);
\draw[arrow,dashed] (D) to[bend right=20] (H);
\draw[arrow,dashed] (D) to[bend right=20] (H);
\draw[arrow,dashed] (E) to[bend right=20] (I);

\draw[arrow,dashed] (H) to[bend right=20] (J);
\draw[arrow,dashed] (H) to[bend right=20] (K);
\draw[arrow,dashed] (I) to[bend right=20] (L);

\begin{scope}[on background layer]
    \draw[red!80, thick, dashed, rounded corners=15pt]
        ([xshift=-5mm,yshift=-2mm]J.south west) -- 
        ([xshift=-5mm]D.west) -- 
        ([xshift=-10mm]B.west) -- 
        ([xshift=0mm,yshift=4mm]A.north west) --
        ([xshift=12mm,yshift=4mm]A.north east) --
        ([xshift=10mm]B.east) --
        ([xshift=10mm]E.east) --
        ([xshift=2mm,yshift=-5mm]H.south east) --
        ([xshift=2mm,yshift=-5mm]J.south east) -- cycle;
\end{scope}

\node[label, red!80!black, text width=2.5cm, align=center] at ([xshift=-2.2cm]B.west) {Tasks to be Reprocessed};

\end{tikzpicture}}
\caption{
    Outcome of traversal phase for an example task tree.
}
\label{fig:corruption_frontier}
\end{figure}
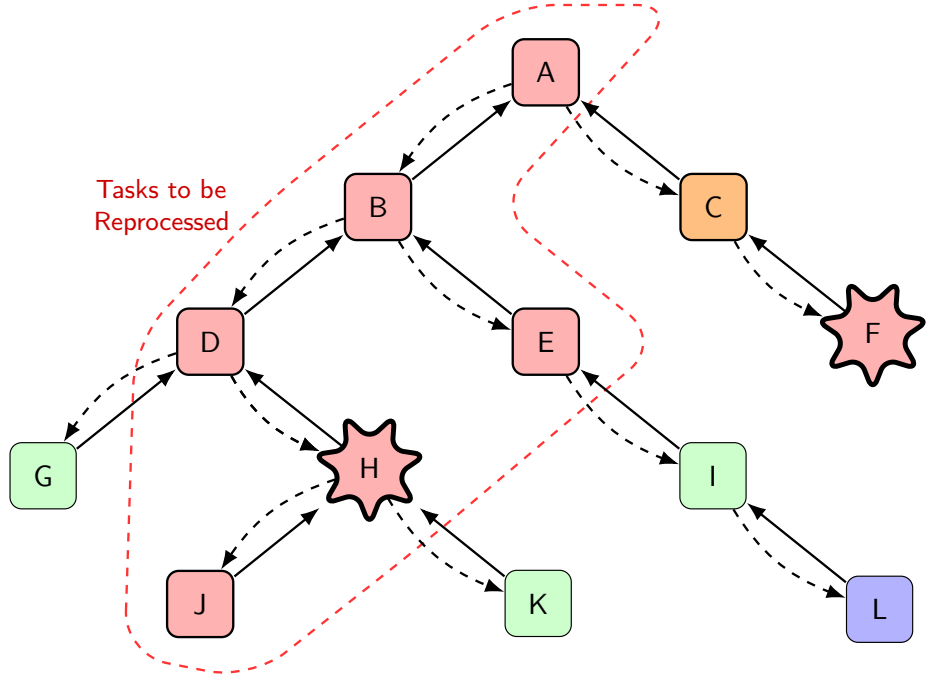

%% file: 05experiments.tex
\section{Experiments}\label{sec:experiments}

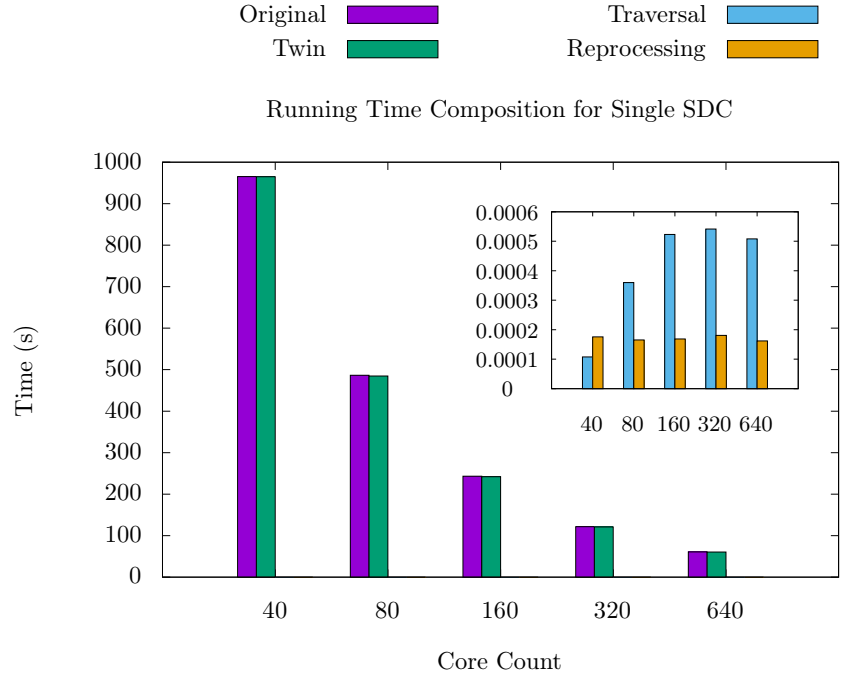
\begin{figure}[t]
    \centering
    \resizebox{1.0\columnwidth}{!}{
      \input{05_fig2}
    }
    \caption{
      Running times of the \textit{original} and \textit{twin} computations (main part), as well as of the traversal and reprocessing phases (inset graph) of recovery for a single SDC.
    }
    \label{fig:composition1SDC}
\end{figure}

\begin{figure}[t]
    \centering
    \resizebox{0.9\columnwidth}{!}{
      \input{05_fig3}
    }
    \caption{
      Duration of the traversal phase as a function of the number of injected SDCs, depicted for different worker counts.
    }
    \label{fig:frontierduration}
\end{figure}
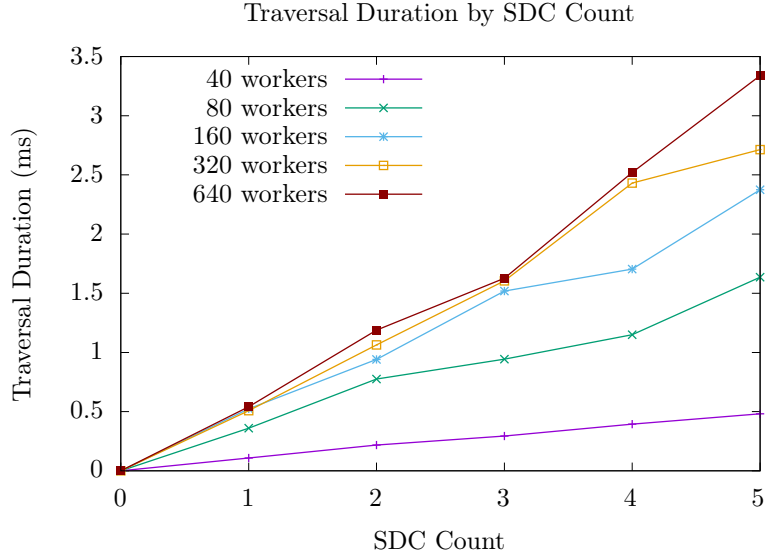

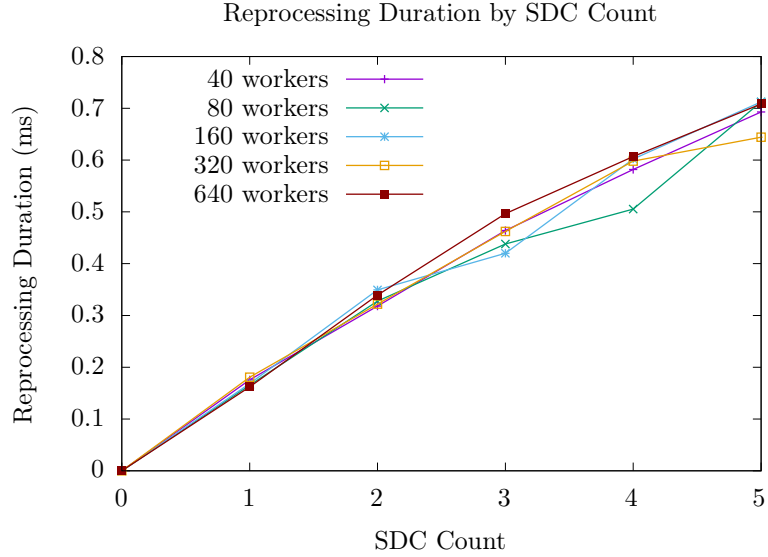
\begin{figure}[t]
    \centering
    \resizebox{0.9\columnwidth}{!}{
      \input{05_fig4}
    }
    \caption{
      Duration of the reprocessing phase as a function of the number of injected SDCs, shown for different worker counts.
    }
    \label{fig:reprocessingduration}
\end{figure}

\subsection{Experimental Setup}

\textbf{Hardware.} All experiments were performed on a partition of the Goethe cluster of the University of Frankfurt~\cite{ClusterGoethe}.
This cluster consists of homogeneous Infiniband-connected nodes, each equipped with two 20-core Intel Xeon Skylake Gold 6148 CPUs and 192~GB of main memory.
We ran our experiments on up to~32 nodes with a total of up to 640~workers.

\textbf{Benchmark.} We adapted the naive recursive Fibonacci~(FIB) benchmark from Section~\ref{sec:background} and set $n = 62$.
To increase the task size, we added a sequential cutoff at ${C=32}$.
This means that all calls to \verb|fib(n)| with $n \leq 32$ are computed sequentially without spawning new tasks.

\textbf{Software.} We compiled our programs with Open~MPI~5.0.5 and g++~11.4.1 using the \texttt{-O3} switch.

\textbf{Fault Injection Methodology.} We injected SDCs into one or several randomly selected tasks of the original computation by flipping random bits of the result right before its return.
The twin computation was not infected.

Each experimental configuration was repeated 20 times, and we report the averages of these runs.

\subsection{Results and Analysis}

Figure~\ref{fig:composition1SDC} depicts  running times for a single SDC across different worker counts.
The primary takeaway is the negligible overhead of the traversal and reprocessing phases.
The time spent in these phases was several orders of magnitude smaller than the execution time of the original and twin computations.
This low overhead confirms the prediction from our analysis in Section~\ref{subsec:analysis}.

Figure~\ref{fig:frontierduration} details the performance of the traversal algorithm as the number of injected SDCs increases from one to five.
The duration scales about linearly with the number of corruptions.
This is expected, as each SDC increases the number of tasks to be visited during traversal.
We also observe that the duration increases with the number of workers.
This is a consequence of our sequential implementation of the traversal phase.
The traversal runs on a single worker, which must retrieve the dependencies from the other workers over the network using MPI one-sided communication.
As the worker count grows, the proportion of dependencies that are stored remotely increases, leading to higher data transfer costs.
However, this increase becomes less pronounced at higher worker counts.
Therefore, doubling the number of workers from 320 to 640 adds proportionally less overhead than doubling it from 40 to 80.

Figure~\ref{fig:reprocessingduration} shows the time required for the reprocessing phase.
The duration increases with the number of SDCs, but notably, the rate of increase slows as more SDCs occur.
This sub-linear scaling is an expected and favorable outcome of our approach.
For programs in which the corruption of a parent task is unlikely to infect its children, an SDC at a given task only invalidates the dependency path to the root.
Thus, when multiple SDCs occur at random locations in the task tree, their respective dependency paths to the root share common tasks.

%% file: 05_fig2.tex
\begingroup
  \makeatletter
  \providecommand\color[2][]{%
    \GenericError{(gnuplot) \space\space\space\@spaces}{%
      Package color not loaded in conjunction with
      terminal option `colourtext'%
    }{See the gnuplot documentation for explanation.%
    }{Either use 'blacktext' in gnuplot or load the package
      color.sty in LaTeX.}%
    \renewcommand\color[2][]{}%
  }%
  \providecommand\includegraphics[2][]{%
    \GenericError{(gnuplot) \space\space\space\@spaces}{%
      Package graphicx or graphics not loaded%
    }{See the gnuplot documentation for explanation.%
    }{The gnuplot epslatex terminal needs graphicx.sty or graphics.sty.}%
    \renewcommand\includegraphics[2][]{}%
  }%
  \providecommand\rotatebox[2]{#2}%
  \@ifundefined{ifGPcolor}{%
    \newif\ifGPcolor
    \GPcolortrue
  }{}%
  \@ifundefined{ifGPblacktext}{%
    \newif\ifGPblacktext
    \GPblacktextfalse
  }{}%
  \let\gplgaddtomacro\g@addto@macro
  \gdef\gplbacktext{}%
  \gdef\gplfronttext{}%
  \makeatother
  \ifGPblacktext
    \def\colorrgb#1{}%
    \def\colorgray#1{}%
  \else
    \ifGPcolor
      \def\colorrgb#1{\color[rgb]{#1}}%
      \def\colorgray#1{\color[gray]{#1}}%
      \expandafter\def\csname LTw\endcsname{\color{white}}%
      \expandafter\def\csname LTb\endcsname{\color{black}}%
      \expandafter\def\csname LTa\endcsname{\color{black}}%
      \expandafter\def\csname LT0\endcsname{\color[rgb]{1,0,0}}%
      \expandafter\def\csname LT1\endcsname{\color[rgb]{0,1,0}}%
      \expandafter\def\csname LT2\endcsname{\color[rgb]{0,0,1}}%
      \expandafter\def\csname LT3\endcsname{\color[rgb]{1,0,1}}%
      \expandafter\def\csname LT4\endcsname{\color[rgb]{0,1,1}}%
      \expandafter\def\csname LT5\endcsname{\color[rgb]{1,1,0}}%
      \expandafter\def\csname LT6\endcsname{\color[rgb]{0,0,0}}%
      \expandafter\def\csname LT7\endcsname{\color[rgb]{1,0.3,0}}%
      \expandafter\def\csname LT8\endcsname{\color[rgb]{0.5,0.5,0.5}}%
    \else
      \def\colorrgb#1{\color{black}}%
      \def\colorgray#1{\color[gray]{#1}}%
      \expandafter\def\csname LTw\endcsname{\color{white}}%
      \expandafter\def\csname LTb\endcsname{\color{black}}%
      \expandafter\def\csname LTa\endcsname{\color{black}}%
      \expandafter\def\csname LT0\endcsname{\color{black}}%
      \expandafter\def\csname LT1\endcsname{\color{black}}%
      \expandafter\def\csname LT2\endcsname{\color{black}}%
      \expandafter\def\csname LT3\endcsname{\color{black}}%
      \expandafter\def\csname LT4\endcsname{\color{black}}%
      \expandafter\def\csname LT5\endcsname{\color{black}}%
      \expandafter\def\csname LT6\endcsname{\color{black}}%
      \expandafter\def\csname LT7\endcsname{\color{black}}%
      \expandafter\def\csname LT8\endcsname{\color{black}}%
    \fi
  \fi
    \setlength{\unitlength}{0.0500bp}%
    \ifx\gptboxheight\undefined%
      \newlength{\gptboxheight}%
      \newlength{\gptboxwidth}%
      \newsavebox{\gptboxtext}%
    \fi%
    \setlength{\fboxrule}{0.5pt}%
    \setlength{\fboxsep}{1pt}%
    \definecolor{tbcol}{rgb}{1,1,1}%
\begin{picture}(7370.00,5668.00)%
    \gplgaddtomacro\gplbacktext{%
      \csname LTb\endcsname
      \put(1204,896){\makebox(0,0)[r]{\strut{}$0$}}%
      \put(1204,1233){\makebox(0,0)[r]{\strut{}$100$}}%
      \put(1204,1570){\makebox(0,0)[r]{\strut{}$200$}}%
      \put(1204,1907){\makebox(0,0)[r]{\strut{}$300$}}%
      \put(1204,2244){\makebox(0,0)[r]{\strut{}$400$}}%
      \put(1204,2582){\makebox(0,0)[r]{\strut{}$500$}}%
      \put(1204,2919){\makebox(0,0)[r]{\strut{}$600$}}%
      \put(1204,3256){\makebox(0,0)[r]{\strut{}$700$}}%
      \put(1204,3593){\makebox(0,0)[r]{\strut{}$800$}}%
      \put(1204,3930){\makebox(0,0)[r]{\strut{}$900$}}%
      \put(1204,4267){\makebox(0,0)[r]{\strut{}$1000$}}%
      \put(2288,616){\makebox(0,0){\strut{}40}}%
      \put(3203,616){\makebox(0,0){\strut{}80}}%
      \put(4119,616){\makebox(0,0){\strut{}160}}%
      \put(5034,616){\makebox(0,0){\strut{}320}}%
      \put(5950,616){\makebox(0,0){\strut{}640}}%
    }%
    \gplgaddtomacro\gplfronttext{%
      \csname LTb\endcsname
      \put(2707,5465){\makebox(0,0)[r]{\strut{}Original}}%
      \csname LTb\endcsname
      \put(2707,5185){\makebox(0,0)[r]{\strut{}Twin}}%
      \csname LTb\endcsname
      \put(5794,5465){\makebox(0,0)[r]{\strut{}Traversal}}%
      \csname LTb\endcsname
      \put(5794,5185){\makebox(0,0)[r]{\strut{}Reprocessing}}%
      \csname LTb\endcsname
      \put(266,2581){\rotatebox{-270.00}{\makebox(0,0){\strut{}Time (s)}}}%
      \put(4118,196){\makebox(0,0){\strut{}Core Count}}%
      \put(4118,4687){\makebox(0,0){\strut{}Running Time Composition for Single SDC}}%
    }%
    \gplgaddtomacro\gplbacktext{%
      \csname LTb\endcsname
      \put(4181,2427){\makebox(0,0){\footnotesize 0}}%
      \put(4181,2667){\makebox(0,0){\footnotesize 0.0001}}%
      \put(4181,2906){\makebox(0,0){\footnotesize 0.0002}}%
      \put(4181,3146){\makebox(0,0){\footnotesize 0.0003}}%
      \put(4181,3385){\makebox(0,0){\footnotesize 0.0004}}%
      \put(4181,3625){\makebox(0,0){\footnotesize 0.0005}}%
      \put(4181,3864){\makebox(0,0){\footnotesize 0.0006}}%
      \put(4867,2147){\makebox(0,0){\footnotesize 40}}%
      \put(5201,2147){\makebox(0,0){\footnotesize 80}}%
      \put(5535,2147){\makebox(0,0){\footnotesize 160}}%
      \put(5869,2147){\makebox(0,0){\footnotesize 320}}%
      \put(6203,2147){\makebox(0,0){\footnotesize 640}}%
    }%
    \gplgaddtomacro\gplfronttext{%
      \csname LTb\endcsname
      \put(2895,3145){\rotatebox{-270.00}{\makebox(0,0){\strut{}}}}%
      \put(5535,1783){\makebox(0,0){\strut{}}}%
      \put(5535,8551){\makebox(0,0){\strut{}}}%
    }%
    \gplbacktext
    \put(0,0){\includegraphics[width={368.50bp},height={283.40bp}]{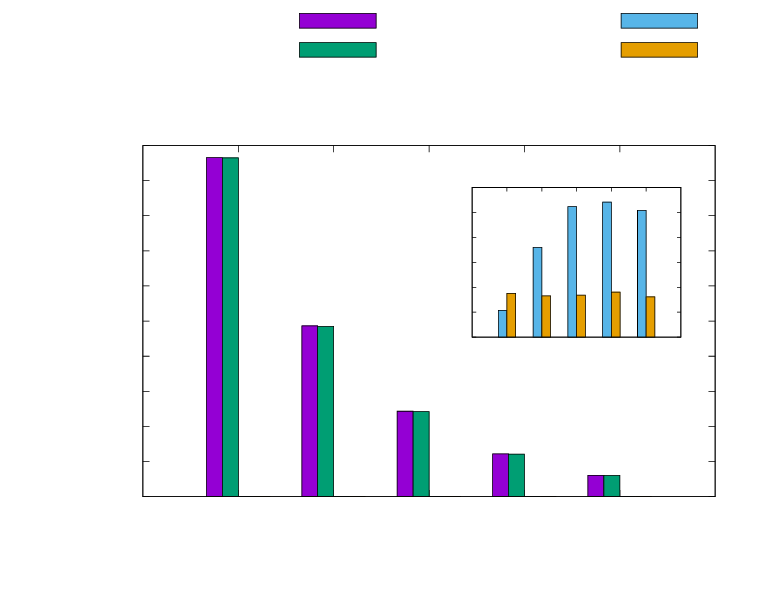}}%
    \gplfronttext
  \end{picture}%
\endgroup

%% file: 05_fig3.tex
\begingroup
  \makeatletter
  \providecommand\color[2][]{%
    \GenericError{(gnuplot) \space\space\space\@spaces}{%
      Package color not loaded in conjunction with
      terminal option `colourtext'%
    }{See the gnuplot documentation for explanation.%
    }{Either use 'blacktext' in gnuplot or load the package
      color.sty in LaTeX.}%
    \renewcommand\color[2][]{}%
  }%
  \providecommand\includegraphics[2][]{%
    \GenericError{(gnuplot) \space\space\space\@spaces}{%
      Package graphicx or graphics not loaded%
    }{See the gnuplot documentation for explanation.%
    }{The gnuplot epslatex terminal needs graphicx.sty or graphics.sty.}%
    \renewcommand\includegraphics[2][]{}%
  }%
  \providecommand\rotatebox[2]{#2}%
  \@ifundefined{ifGPcolor}{%
    \newif\ifGPcolor
    \GPcolortrue
  }{}%
  \@ifundefined{ifGPblacktext}{%
    \newif\ifGPblacktext
    \GPblacktextfalse
  }{}%
  \let\gplgaddtomacro\g@addto@macro
  \gdef\gplbacktext{}%
  \gdef\gplfronttext{}%
  \makeatother
  \ifGPblacktext
    \def\colorrgb#1{}%
    \def\colorgray#1{}%
  \else
    \ifGPcolor
      \def\colorrgb#1{\color[rgb]{#1}}%
      \def\colorgray#1{\color[gray]{#1}}%
      \expandafter\def\csname LTw\endcsname{\color{white}}%
      \expandafter\def\csname LTb\endcsname{\color{black}}%
      \expandafter\def\csname LTa\endcsname{\color{black}}%
      \expandafter\def\csname LT0\endcsname{\color[rgb]{1,0,0}}%
      \expandafter\def\csname LT1\endcsname{\color[rgb]{0,1,0}}%
      \expandafter\def\csname LT2\endcsname{\color[rgb]{0,0,1}}%
      \expandafter\def\csname LT3\endcsname{\color[rgb]{1,0,1}}%
      \expandafter\def\csname LT4\endcsname{\color[rgb]{0,1,1}}%
      \expandafter\def\csname LT5\endcsname{\color[rgb]{1,1,0}}%
      \expandafter\def\csname LT6\endcsname{\color[rgb]{0,0,0}}%
      \expandafter\def\csname LT7\endcsname{\color[rgb]{1,0.3,0}}%
      \expandafter\def\csname LT8\endcsname{\color[rgb]{0.5,0.5,0.5}}%
    \else
      \def\colorrgb#1{\color{black}}%
      \def\colorgray#1{\color[gray]{#1}}%
      \expandafter\def\csname LTw\endcsname{\color{white}}%
      \expandafter\def\csname LTb\endcsname{\color{black}}%
      \expandafter\def\csname LTa\endcsname{\color{black}}%
      \expandafter\def\csname LT0\endcsname{\color{black}}%
      \expandafter\def\csname LT1\endcsname{\color{black}}%
      \expandafter\def\csname LT2\endcsname{\color{black}}%
      \expandafter\def\csname LT3\endcsname{\color{black}}%
      \expandafter\def\csname LT4\endcsname{\color{black}}%
      \expandafter\def\csname LT5\endcsname{\color{black}}%
      \expandafter\def\csname LT6\endcsname{\color{black}}%
      \expandafter\def\csname LT7\endcsname{\color{black}}%
      \expandafter\def\csname LT8\endcsname{\color{black}}%
    \fi
  \fi
    \setlength{\unitlength}{0.0500bp}%
    \ifx\gptboxheight\undefined%
      \newlength{\gptboxheight}%
      \newlength{\gptboxwidth}%
      \newsavebox{\gptboxtext}%
    \fi%
    \setlength{\fboxrule}{0.5pt}%
    \setlength{\fboxsep}{1pt}%
    \definecolor{tbcol}{rgb}{1,1,1}%
\begin{picture}(6236.00,4534.00)%
    \gplgaddtomacro\gplbacktext{%
      \csname LTb\endcsname
      \put(814,704){\makebox(0,0)[r]{\strut{}$0$}}%
      \put(814,1157){\makebox(0,0)[r]{\strut{}$0.5$}}%
      \put(814,1609){\makebox(0,0)[r]{\strut{}$1$}}%
      \put(814,2062){\makebox(0,0)[r]{\strut{}$1.5$}}%
      \put(814,2515){\makebox(0,0)[r]{\strut{}$2$}}%
      \put(814,2968){\makebox(0,0)[r]{\strut{}$2.5$}}%
      \put(814,3420){\makebox(0,0)[r]{\strut{}$3$}}%
      \put(814,3873){\makebox(0,0)[r]{\strut{}$3.5$}}%
      \put(946,484){\makebox(0,0){\strut{}$0$}}%
      \put(1925,484){\makebox(0,0){\strut{}$1$}}%
      \put(2903,484){\makebox(0,0){\strut{}$2$}}%
      \put(3882,484){\makebox(0,0){\strut{}$3$}}%
      \put(4860,484){\makebox(0,0){\strut{}$4$}}%
      \put(5839,484){\makebox(0,0){\strut{}$5$}}%
    }%
    \gplgaddtomacro\gplfronttext{%
      \csname LTb\endcsname
      \put(2530,3700){\makebox(0,0)[r]{\strut{}40 workers}}%
      \csname LTb\endcsname
      \put(2530,3480){\makebox(0,0)[r]{\strut{}80 workers}}%
      \csname LTb\endcsname
      \put(2530,3260){\makebox(0,0)[r]{\strut{}160 workers}}%
      \csname LTb\endcsname
      \put(2530,3040){\makebox(0,0)[r]{\strut{}320 workers}}%
      \csname LTb\endcsname
      \put(2530,2820){\makebox(0,0)[r]{\strut{}640 workers}}%
      \csname LTb\endcsname
      \put(209,2288){\rotatebox{-270.00}{\makebox(0,0){\strut{}Traversal Duration (ms)}}}%
      \put(3392,154){\makebox(0,0){\strut{}SDC Count}}%
      \put(3392,4203){\makebox(0,0){\strut{}Traversal Duration by SDC Count}}%
    }%
    \gplbacktext
    \put(0,0){\includegraphics[width={311.80bp},height={226.70bp}]{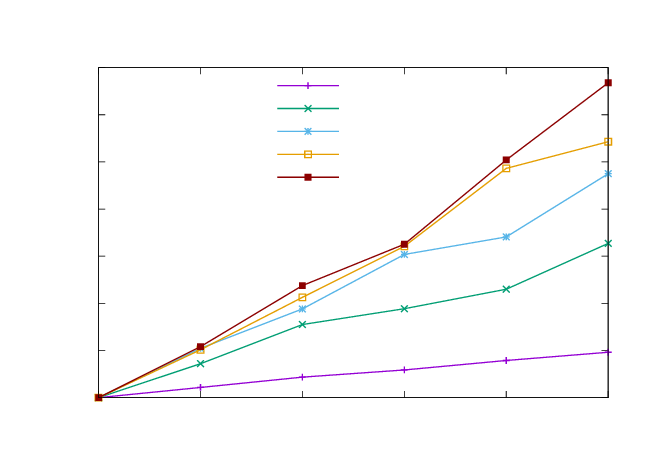}}%
    \gplfronttext
  \end{picture}%
\endgroup

%% file: 05_fig4.tex
\begingroup
  \makeatletter
  \providecommand\color[2][]{%
    \GenericError{(gnuplot) \space\space\space\@spaces}{%
      Package color not loaded in conjunction with
      terminal option `colourtext'%
    }{See the gnuplot documentation for explanation.%
    }{Either use 'blacktext' in gnuplot or load the package
      color.sty in LaTeX.}%
    \renewcommand\color[2][]{}%
  }%
  \providecommand\includegraphics[2][]{%
    \GenericError{(gnuplot) \space\space\space\@spaces}{%
      Package graphicx or graphics not loaded%
    }{See the gnuplot documentation for explanation.%
    }{The gnuplot epslatex terminal needs graphicx.sty or graphics.sty.}%
    \renewcommand\includegraphics[2][]{}%
  }%
  \providecommand\rotatebox[2]{#2}%
  \@ifundefined{ifGPcolor}{%
    \newif\ifGPcolor
    \GPcolortrue
  }{}%
  \@ifundefined{ifGPblacktext}{%
    \newif\ifGPblacktext
    \GPblacktextfalse
  }{}%
  \let\gplgaddtomacro\g@addto@macro
  \gdef\gplbacktext{}%
  \gdef\gplfronttext{}%
  \makeatother
  \ifGPblacktext
    \def\colorrgb#1{}%
    \def\colorgray#1{}%
  \else
    \ifGPcolor
      \def\colorrgb#1{\color[rgb]{#1}}%
      \def\colorgray#1{\color[gray]{#1}}%
      \expandafter\def\csname LTw\endcsname{\color{white}}%
      \expandafter\def\csname LTb\endcsname{\color{black}}%
      \expandafter\def\csname LTa\endcsname{\color{black}}%
      \expandafter\def\csname LT0\endcsname{\color[rgb]{1,0,0}}%
      \expandafter\def\csname LT1\endcsname{\color[rgb]{0,1,0}}%
      \expandafter\def\csname LT2\endcsname{\color[rgb]{0,0,1}}%
      \expandafter\def\csname LT3\endcsname{\color[rgb]{1,0,1}}%
      \expandafter\def\csname LT4\endcsname{\color[rgb]{0,1,1}}%
      \expandafter\def\csname LT5\endcsname{\color[rgb]{1,1,0}}%
      \expandafter\def\csname LT6\endcsname{\color[rgb]{0,0,0}}%
      \expandafter\def\csname LT7\endcsname{\color[rgb]{1,0.3,0}}%
      \expandafter\def\csname LT8\endcsname{\color[rgb]{0.5,0.5,0.5}}%
    \else
      \def\colorrgb#1{\color{black}}%
      \def\colorgray#1{\color[gray]{#1}}%
      \expandafter\def\csname LTw\endcsname{\color{white}}%
      \expandafter\def\csname LTb\endcsname{\color{black}}%
      \expandafter\def\csname LTa\endcsname{\color{black}}%
      \expandafter\def\csname LT0\endcsname{\color{black}}%
      \expandafter\def\csname LT1\endcsname{\color{black}}%
      \expandafter\def\csname LT2\endcsname{\color{black}}%
      \expandafter\def\csname LT3\endcsname{\color{black}}%
      \expandafter\def\csname LT4\endcsname{\color{black}}%
      \expandafter\def\csname LT5\endcsname{\color{black}}%
      \expandafter\def\csname LT6\endcsname{\color{black}}%
      \expandafter\def\csname LT7\endcsname{\color{black}}%
      \expandafter\def\csname LT8\endcsname{\color{black}}%
    \fi
  \fi
    \setlength{\unitlength}{0.0500bp}%
    \ifx\gptboxheight\undefined%
      \newlength{\gptboxheight}%
      \newlength{\gptboxwidth}%
      \newsavebox{\gptboxtext}%
    \fi%
    \setlength{\fboxrule}{0.5pt}%
    \setlength{\fboxsep}{1pt}%
    \definecolor{tbcol}{rgb}{1,1,1}%
\begin{picture}(6236.00,4534.00)%
    \gplgaddtomacro\gplbacktext{%
      \csname LTb\endcsname
      \put(814,704){\makebox(0,0)[r]{\strut{}$0$}}%
      \put(814,1100){\makebox(0,0)[r]{\strut{}$0.1$}}%
      \put(814,1496){\makebox(0,0)[r]{\strut{}$0.2$}}%
      \put(814,1892){\makebox(0,0)[r]{\strut{}$0.3$}}%
      \put(814,2289){\makebox(0,0)[r]{\strut{}$0.4$}}%
      \put(814,2685){\makebox(0,0)[r]{\strut{}$0.5$}}%
      \put(814,3081){\makebox(0,0)[r]{\strut{}$0.6$}}%
      \put(814,3477){\makebox(0,0)[r]{\strut{}$0.7$}}%
      \put(814,3873){\makebox(0,0)[r]{\strut{}$0.8$}}%
      \put(946,484){\makebox(0,0){\strut{}$0$}}%
      \put(1925,484){\makebox(0,0){\strut{}$1$}}%
      \put(2903,484){\makebox(0,0){\strut{}$2$}}%
      \put(3882,484){\makebox(0,0){\strut{}$3$}}%
      \put(4860,484){\makebox(0,0){\strut{}$4$}}%
      \put(5839,484){\makebox(0,0){\strut{}$5$}}%
    }%
    \gplgaddtomacro\gplfronttext{%
      \csname LTb\endcsname
      \put(2530,3700){\makebox(0,0)[r]{\strut{}40 workers}}%
      \csname LTb\endcsname
      \put(2530,3480){\makebox(0,0)[r]{\strut{}80 workers}}%
      \csname LTb\endcsname
      \put(2530,3260){\makebox(0,0)[r]{\strut{}160 workers}}%
      \csname LTb\endcsname
      \put(2530,3040){\makebox(0,0)[r]{\strut{}320 workers}}%
      \csname LTb\endcsname
      \put(2530,2820){\makebox(0,0)[r]{\strut{}640 workers}}%
      \csname LTb\endcsname
      \put(209,2288){\rotatebox{-270.00}{\makebox(0,0){\strut{}Reprocessing Duration (ms)}}}%
      \put(3392,154){\makebox(0,0){\strut{}SDC Count}}%
      \put(3392,4203){\makebox(0,0){\strut{}Reprocessing Duration by SDC Count}}%
    }%
    \gplbacktext
    \put(0,0){\includegraphics[width={311.80bp},height={226.70bp}]{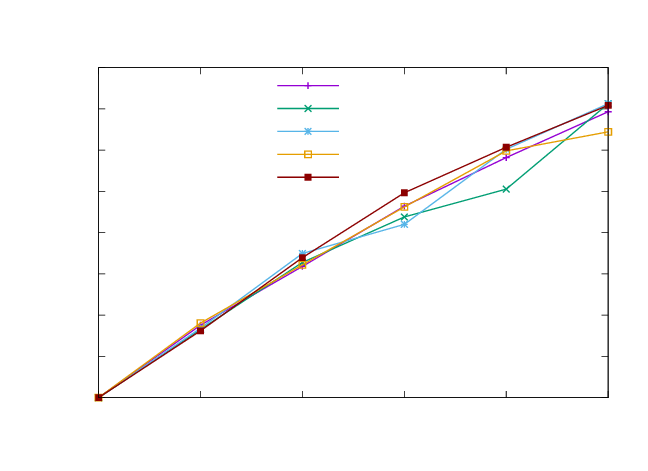}}%
    \gplfronttext
  \end{picture}%
\endgroup

%% file: 06extension.tex
\section{Extension to Future-Based Cooperation}\label{sec:extension}

\input{06_fig5.tex}

So far, we have assumed that futures are only touched by the parent of their filling task.
In a more relaxed task model, called Future-based Cooperation~(FBC)~\cite{MiaFutureSideEffects}, tasks may pass futures as arguments to child tasks and tasks may return futures that contain futures (e.g., futures of their child task results).
This gives rise to dependencies between tasks that are not in a parent-child relationship, and thus to a task dependency structure that is a Directed Acyclic Graph~(DAG) instead of a tree.
Figure~\ref{fig:fbcissue} shows an example, in which the results of~B and~E are not used by their parent, but by task~C.

Extending our scheme from Section~\ref{sec:design} to the FBC setting is not trivial.
In the example of Figure~\ref{fig:fbcissue}, for instance, task~B would not be re-spawn during the reprocessing phase, and thus~E could not be re-invoked.
Moreover, the corrupted task D would not be recognized as such, despite a possible impact on the final result.

The problem can be overcome by modifying the traversal algorithm such that, upon identifying a corrupted task, all of its ancestors are marked corrupted as well, and then the traversal is applied to all of their descendants.
Details are left for future work.

%% file: 06_fig5.tex
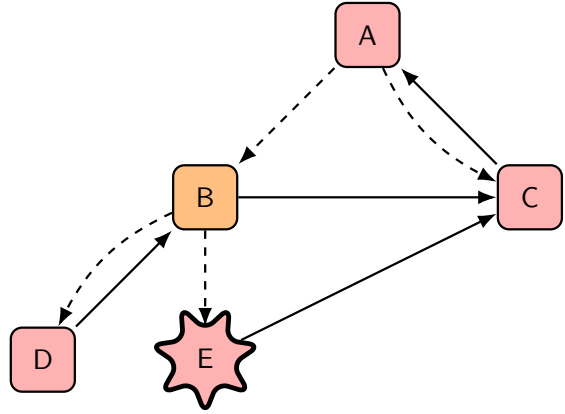
\begin{figure}[h!]
\centering
\resizebox{0.6\columnwidth}{!}{\begin{tikzpicture}[
    node distance=1.2cm and 1.2cm,
    task/.style={
        draw,
        rectangle,
        rounded corners,
        minimum size=8mm,
        font=\sffamily
    },
    corrupted/.style={
        task,
        fill=red!30,
        thick
    },
    correct/.style={
        task,
        fill=green!20
    },
    frontier/.style={
        task,
        fill=blue!30,
        thick,
        line width=0.6mm
    },
    infected/.style={
        task,
        fill=orange!50,
        thick
    },
    sdc_source/.style={
        corrupted,
        star,
        star points=7,
        star point ratio=0.5,
        minimum size=7mm,
        line width=0.6mm
    },
    arrow/.style={
        -Latex,
        thick
    },
    label/.style={
        font=\sffamily\small
    }
]

\node[corrupted] (A) {A};

\node[infected, below left=of A] (B) {B};
\node[corrupted, below right=of A] (C) {C};
\node[corrupted, below left=of B] (D) {D};

\node[sdc_source, below=of B] (E) {E};

\draw[arrow, dashed] (A) to[bend right=20] (C);
\draw[arrow] (C) to  (A);
\draw[arrow,dashed] (A) -- (B);
\draw[arrow] (B) -- (C);
\draw[arrow,dashed] (B) to[bend right=20] (D);
\draw[arrow] (D) -- (B);

\draw[arrow,dashed] (B) -- (E);
\draw[arrow] (E) -- (C);

\end{tikzpicture}}
\caption{
    Example DAG for FBC.
}
\label{fig:fbcissue}
\end{figure}

%% file: 07conclusions.tex
\section{Conclusions}\label{sec:conclusions}

This paper has introduced a novel replication-based scheme to provide efficient SDC protection for NFJ programs.

Our approach runs two instances of a computation, records their task dependencies, only if the final results differ, performs a post-mortem analysis of the task trees, and afterwards re-runs the potentially affected parts of the computation.

Our implementation and evaluation within the ItoyoriFBC runtime have demonstrated the correctness and efficiency of the new approach.
In line with a theoretical analysis performed in this paper, we observed a negligible overhead for the traversal and reprocessing phases.

For future work, several avenues exist.
Most importantly, the discussed adaptation for FBC from Section~\ref{sec:extension} should be investigated, and
the implementation of the reprocessing phase, which uses a single worker, should be parallelized.

%% file: bibo.bib
@article{ABFTOrig,
    author = {Kuang-Hua Huang and Jacob A. Abraham},
    title = {Algorithm-Based Fault Tolerance for Matrix Operations},
    year = {1984},
    publisher = {IEEE},
    volume = {33},
    number = {6},
    doi = {10.1109/TC.1984.1676475},
    journal = {Transaction on Computers},
    pages = {518–528},
}

@online{ClusterGoethe,
    author = {TOP500.org},
    title = {{Goethe-NHR}},
    urldate = {2025-06-23},
    url = {https://www.top500.org/system/179588},
}

@book{FTBuch,
    editor = {Thomas Herault and Yves Robert},
    publisher = {Springer},
    title = {Fault-Tolerance Techniques for High-Performance Computing},
    year = {2015},
    doi = {10.1007/978-3-319-20943-2},
}

@article{ABFTBos,
    author = {George Bosilca and R{\'{e}}mi Delmas and Jack Dongarra and Julien Langou},
    journal = {Journal of Parallel and Distributed Computing (JPDC)},
    number = {4},
    pages = {410-416},
    publisher = {Elsevier},
    title = {Algorithm-based fault tolerance applied to high performance computing},
    volume = {69},
    year = {2009},
    doi = {10.1016/j.jpdc.2008.12.002},
}

@article{ABFT,
    author = {Nawab Ali and Sriram Krishnamoorthy and Mahantesh Halappanavar and Jeff Daily},
    journal = {International Journal of Parallel Programming (JPDC)},
    number = {3},
    pages = {469-493},
    publisher = {Springer},
    title = {Multi-Fault Tolerance for Cartesian Data Distributions},
    volume = {41},
    year = {2012},
    doi = {10.1007/s10766-012-0218-5},
}

@inproceedings{SoftKrishna,
    author = {Mehmet Can Kurt and Sriram Krishnamoorthy and Kunal Agrawal and Gagan Agrawal},
    booktitle = {Proceedings International Conference for High Performance Computing, Networking, Storage and Analysis (SC)},
    pages = {719-730},
    publisher = {ACM},
    title = {Fault-Tolerant Dynamic Task Graph Scheduling},
    year = {2014},
    doi = {10.1109/SC.2014.64},
}

@inproceedings{KrishnaNestedFJournal,
    author = {Gokcen Kestor and Sriram Krishnamoorthy and Wenjing Ma},
    booktitle = {Proceedings International Symposium on Parallel and Distributed Processing (IPDPS)},
    title = {Localized Fault Recovery for Nested Fork-Join Programs},
    pages = {397-408},
    publisher = {IEEE},
    year = {2017},
    doi = {10.1109/ipdps.2017.75},
}

@article{JonasIJNC22,
    author = {Jonas Posner and Mia Reitz and Claudia Fohry},
    title = {Task-Level Resilience: Checkpointing vs. Supervision},
    journal = {Special Issue International Journal of Networking and Computing (IJNC)},
    year = {2022},
    volume = {12},
    number = {1},
    pages = {47-72},
    doi = {10.15803/ijnc.12.1_47},
}

@article{JonasIncFTGLBJournal19,
    author = {Jonas Posner and Mia Reitz and Claudia Fohry},
    title = {A Comparison of Application-Level Fault Tolerance Schemes for Task Pools},
    journal = {Future Generation Computing Systems (FGCS)},
    year = {2019},
    volume = {105},
    pages = {119-134},
    doi = {10.1016/j.future.2019.11.031},
}

@inproceedings{MiaNFJCheckpointing,
    author = {Mia Reitz and Claudia Fohry},
    title = {Task-Level Checkpointing for Nested Fork-Join Programs using Work Stealing},
    booktitle = {Workshop on Asynchronous Many-Task Systems for Exascale (AMTE)},
    year = {2023},
    publisher = {Springer},
    doi = {10.1007/978-3-031-48803-0_9},
}

@inproceedings{DoubtAndRedundancySoft,
author = { Philipp Samfass and Tobias Weinzierl and Anne Reinarz and Michael Bader},
booktitle = { Workshop on Fault Tolerance for {HPC} at {eXtreme} Scale {(FTXS)} },
title = { Doubt and Redundancy Kill Soft Errors—Towards Detection and Correction of Silent Data Corruption in Task-based Numerical Software },
year = {2021},
volume = {},
ISSN = {},
pages = {1-10},
doi = {10.1109/FTXS54580.2021.00005},
}

@inproceedings{MiaFutureSideEffects,
    author = {Mia Reitz and Ben Gerhards and John Hundhausen and Claudia Fohry},
    title = {Investigating the Performance Difference of Task Communication via Futures or Side Effects},
    booktitle = {Workshop on Asynchronous Many-Task Systems for Exascale (AMTE)},
    year = {2024},
    publisher = {Springer},
    doi = {10.1007/978-3-031-90200-0_21},
}

@inproceedings{MiaFuturesCheckpointing,
    author = {Mia Reitz and John Hundhausen and Claudia Fohry},
    title = {Fail-stop Failure Protection for Coordinated Work Stealing of Tasks that Communicate Through Futures},
    booktitle = {Workshop on Asynchronous Many-Task Systems and Applications (WAMTA)},
    year = {2025},
    doi = {},
}

@inproceedings{itoyori,
    author =        {Shumpei Shiina and Kenjiro Taura},
    booktitle =     {Proceedings of the International Conference for High
                   Performance Computing, Networking, Storage and
                   Analytics},
    title =         {Itoyori: Reconciling Global Address Space and Global
                   Fork-Join Task Parallelism},
    year =          {2023},
    doi =           {10.1145/3581784.3607049},
}

@INPROCEEDINGS{ftx1,
    author = {Hemanth Kolla and Jackson R. Mayo and Keita Teranishi and Robert C. Armstrong},
    title = {Improving Scalability of Silent-Error Resilience for Message-Passing Solvers via Local Recovery and Asynchrony},
    booktitle = {Proceedings Workshop on Fault Tolerance for HPC at eXtreme Scale (FTXS)},
    year = {2020},
    pages = {1--10},
    doi = {10.1109/FTXS51974.2020.00006},
}

@article{originalReplication,
  author = {Fred B. Schneider},
  title = {Implementing fault-tolerant services using the state machine approach: a tutorial},
  year = {1990},
  publisher = {Association for Computing Machinery},
  address = {New York, NY, USA},
  volume = {22},
  number = {4},
  issn = {0360-0300},
  doi = {10.1145/98163.98167},
  journal = {ACM Comput. Surv.},
  pages = {299–319},
  numpages = {21}
}

@inproceedings{HOPE,
  author = {Masahiro Yasugi and Daisuke Muraoka and Tasuku Hiraishi and Seiji Umatani and Kento Emoto},
  title = {{HOPE}: A Parallel Execution Model Based on Hierarchical Omission},
  year = {2019},
  isbn = {9781450362955},
  doi = {10.1145/3337821.3337899},
  booktitle = {Proceedings of the 48th International Conference on Parallel Processing},
  articleno = {77},
  numpages = {11},
  series = {ICPP}
}
